\documentclass[conference]{IEEEtran}
\IEEEoverridecommandlockouts
\usepackage{cite}
\usepackage{amsmath,amssymb,amsfonts}
\usepackage{algorithmic}
\usepackage{graphicx}
\usepackage{textcomp}
\usepackage{subfigure}
\usepackage{multirow}
\usepackage{xcolor}
\usepackage{hyperref}
\usepackage{breakurl}

\def\BibTeX{{\rm B\kern-.05em{\sc i\kern-.025em b}\kern-.08em
    T\kern-.1667em\lower.7ex\hbox{E}\kern-.125emX}}

\usepackage{booktabs}
\usepackage{colortbl}

\usepackage{upgreek}
\renewcommand{\mu}{\upmu}

\begin{document}

\title{MPI Errors Detection using GNN Embedding and Vector Embedding over LLVM IR\\
}

\author{\IEEEauthorblockN{Jad El Karchi\textsuperscript{2}, Hanze Chen\textsuperscript{1}, Ali TehraniJamsaz\textsuperscript{1}, Ali Jannesari\textsuperscript{1},  Mihail Popov\textsuperscript{2}, Emmanuelle Saillard\textsuperscript{2}}
\IEEEauthorblockA{\textsuperscript{1}\textit{Iowa State University, Ames, Iowa, USA}}
\IEEEauthorblockA{\textsuperscript{2}\textit{Inria, Bordeaux, France}}
\IEEEauthorblockA{\{hanzech, tehrani, jannesar\}@iastate.edu}
\IEEEauthorblockA{\{jad.el-karchi, mihail.popov, emmanuelle.saillard\}@inria.fr}
\\[-3.0ex]
}

\maketitle

\begin{abstract}

Identifying errors in parallel MPI programs is a challenging task. Despite the growing number of verification tools, debugging parallel programs remains a significant challenge. This paper is the first to utilize embedding and deep learning graph neural networks (GNNs) to tackle the issue of identifying bugs in MPI programs. Specifically, we have designed and developed two models that can determine, from a code's LLVM Intermediate Representation (IR), whether the code is correct or contains a known MPI error.


We tested our models using two dedicated MPI benchmark suites for verification: MBI and MPI-CorrBench. By training and validating our models on the same benchmark suite, we achieved a prediction accuracy of 92\% in detecting error types. Additionally, we trained and evaluated our models on distinct benchmark suites (e.g., transitioning from MBI to MPI-CorrBench) and achieved a promising accuracy of over 80\%. Finally, we investigated the interaction between different MPI errors and quantified our models generalization capabilities over new unseen errors.
This involved removing errors types during training and assessing whether our models could still predict them. The detection accuracy of removed errors vary significantly between 20\% to 80\%, indicating connected error patterns.

\end{abstract}


\section{Introduction}

High-Performance Computing (HPC) plays an important role in many fields like health, materials science, security, and the environment. The current supercomputer hardware trends lead to more complex parallel applications with heterogeneity in hardware, new scalable algorithms, and combinations of parallel programming models that pose many programmability challenges. This demands a requirement for more efficient and scalable debugging techniques to assist HPC application developers and parallel programming. Yet, despite the growing number of verification and debugging tools, determining if a parallel program always behaves as expected on any execution is challenging due to non-deterministic executions~\cite{correctness17}. 

MPI is one of the most popular programming models in High-Performance Computing. In an MPI program, each MPI process executes a parallel instance of a program in a private address space and exchanges data across distributed memory systems via messages. MPI exposes many ways of exchanging data, including collectives, point-to-point, persistent, and one-sided communications: many errors can occur in an MPI program.

This paper proposes a new AI-assisted approach for detecting errors in MPI programs.
In particular, we devise a machine learning approach that leverages the representational learning of programs to identify errors.
The experimental results indicate that the approach is highly effective in identifying different types of errors or assessing if a code is correct.
In summary, this paper makes the following contributions:
\begin{itemize}
    \item ML approaches with MPI errors detection capabilities that are on par with existing verification tools. They only require a labeled code dataset and can be applied to new scenarios. 
    \item A thorough evaluation of the error detection mechanisms over two dedicated benchmark suites: the MPI Bugs Initiative (MBI)~\cite{mbi} and MPI-CorrBench~\cite{corrbench21}.
    \item A quantification of the generalization of our models by predicting across benchmark suites, by removing errors in the training set and looking for them in the validation, and by studying an error in a real application.
    %
\end{itemize}
The rest of the paper is structured as follows:
The next section discusses related works. Section~\ref{sec:dataset} describes the datasets we use for our experiments, while Section~\ref{sec:methods} explains the details of the proposed approach. Section~\ref{sec:results} provides experimental results. Section~\ref{sec:lim} discusses our limitations and future work.

\section{Related Works}

\subsection{MPI Verification Method}

Related works on MPI program verification use many approaches to detect errors in MPI applications, including static analysis, symbolic execution, concolic testing, model checking, dynamic verification techniques, MPI special libraries and trace-based approaches. Because of the large diversity of programming features and complexity of current systems, no existing tool is currently able to detect all possible errors~\cite{mbi,corrbench21}. They all come with restriction: they are focused on one programming model, they target a specific error, they work with a specific MPI implementation, or they are not yet mature.

Static analyses enable an early errors detection (i.e., the program is not executed) but can report false positives (an error is reported on a correct scenario). Among static tools, MPI-Checker~\cite{mpichecker} is based on the Clang Static Analyzer. It performs so-called AST-based and path-sensitive checks. AST-based checks include correct type usage while path-sensitive checks verify aspects of nonblocking communication, based on the usage of MPI requests.
CIVL~\cite{civl} and MPI-SV~\cite{mpi-sv} both combine symbolic execution and model checking to detect communication deadlocks.
MPISE and Hermes~\cite{hermes} detect communication deadlocks with concolic testing. This method, proven efficient on sequential programs, performs symbolic execution dynamically with a concrete execution \cite{cute, dart}. Compared to these two tools, COMPI~\cite{compi18,compi19} uses input tuning to achieve high branch coverage and tackles runtime bugs like assertion violation or infinite loops.
SimGridMC~\cite{simgrid} and ISP~\cite{isp} check if a program satisfies a given property (e.g., liveness, communication determinism) by considering all possible executions. Like all model checkers, they face the state space explosion problem.
Aislin~\cite{aislin} is an explicit-state model checker which verifies MPI programs with arbitrary-sized system buffers.
Dynamic verification techniques, such as MUST~\cite{must} detect runtime errors. These tools find real errors but stand for a specific environment and can miss errors (false negatives). MUST intercepts all MPI operations to perform online checking. It is based on GTI~\cite{hilbrich2012gti} (Generic Tool Infrastructure) and can detect multiple errors like deadlocks, type mismatches or resource leaking.
DAMPI~\cite{dampi} and Intel Trace Analyzer and Collector (ITAC)~\cite{itac} detect deadlocks with a time-out approach.
MC-CChecker~\cite{mccchecker} and MC-Checker~\cite{mcchecker} both use a trace-based approach to detect memory consistency. They focus on MPI Remote Memory Access (RMA) correctness and do not support other MPI features.
Validation can also be done inside MPI libraries such as in MPICH~\cite{mpich2} or NEC-MPI~\cite{necmpi}. However, the detection of errors is limited to the information available to the MPI routines.
PARCOACH~\cite{parcoachJ,parcoach18} combines static analysis with code instrumentation to detect misuse of MPI collectives and data races that can occur when using nonblocking and persistent communications as well as one-sided communications~\cite{parcoach20, aitkaci:hal-03374614, saillard:hal-03882459}. Although it uses a precise data- and control-flow interprocedural analysis to pinpoint root cause problems, it may lead to many false positives.

All tools cited are using their own terms to report errors and are subject to runtime and compilation failure if a feature is not supported by the tool. In this paper, we propose a new method that learns incorrect patterns by studying the source code (i.e., the compiler Intermediate Representation specifically), irrespective of the language or the MPI implementation used. Our method detects all errors present in the benchmark dataset. We also investigate the importance of each MPI error type with an ablation study that removes errors from the learning dataset and tries to detect them during validation.

\subsection{ML for Bug Detection}

Novel ML techniques have emerged for bug detection and code refactoring. They efficiently detect and potentially correct issues without humans costly hand-crafting detectors (e.g., variable misuse \cite{allamanis2017learning}, wrong binary operator \cite{pradel2018deepbugs}, comment deletion in Python~\cite{allamanis2021self}, or specific Javascript functions~\cite{dinella2020hoppity}). 
The insight is to associate patterns in the source code to identified bugs  with Deep Neural Network (DNN)-based models. To check a new code, the DNN just needs to search for these patterns within the source code. While the source code is directly written by the developer, it is not the only used code representation to detect patterns. Intermediate structures manipulated by the compiler such as the Abstract Syntax Tree (AST), a tree representation of the abstract syntactic structure of a source code, or the LLVM Intermediate Representation\cite{lattner2004llvm} (IR), an accurate and language/hardware independent code representation potentially used with a virtual machine, are also used by ML techniques~\cite{white2019sorting,white2016deep,mou2016convolutional,wei2017supervised,venkatakeerthy2020ir2vec}. Nevertheless and to the best of our knowledge, despite this diversity of representations, ML techniques currently fail to detect bugs in parallel programs either because the representations fail to efficiently expose the contexts of the patterns or because of insufficient training data.

\subsection{Vulnerability Detection}
In software developments, vulnerabilities are those security flaws or weaknesses that attackers can exploit.
There have been works that use expert-defined insights~\cite{ayewah2010google} or symbolic execution \cite{cadar2013symbolic, ramos2015under} to identify vulnerabilities. 
Recently, machine learning has also been used for detecting vulnerabilities \cite{li2018vuldeepecker, zou2019mu, lin2019software}. While vulnerability detection has benefited from machine learning-based approaches, MPI error detection, on the other hand, has not received the same attention. One of the reasons is that MPI errors are mostly non-deterministic and hard to identify. In this paper, we aim to fill in this gap by targeting MPI errors and leveraging the latest techniques in machine learning.

To the best of our knowledge, we propose the first method using ML to detect errors specifically in MPI applications.

\begin{figure*}[ht!]
  \centering
  \subfigure[MPI-CorrBench]{\includegraphics[width=0.44\linewidth]{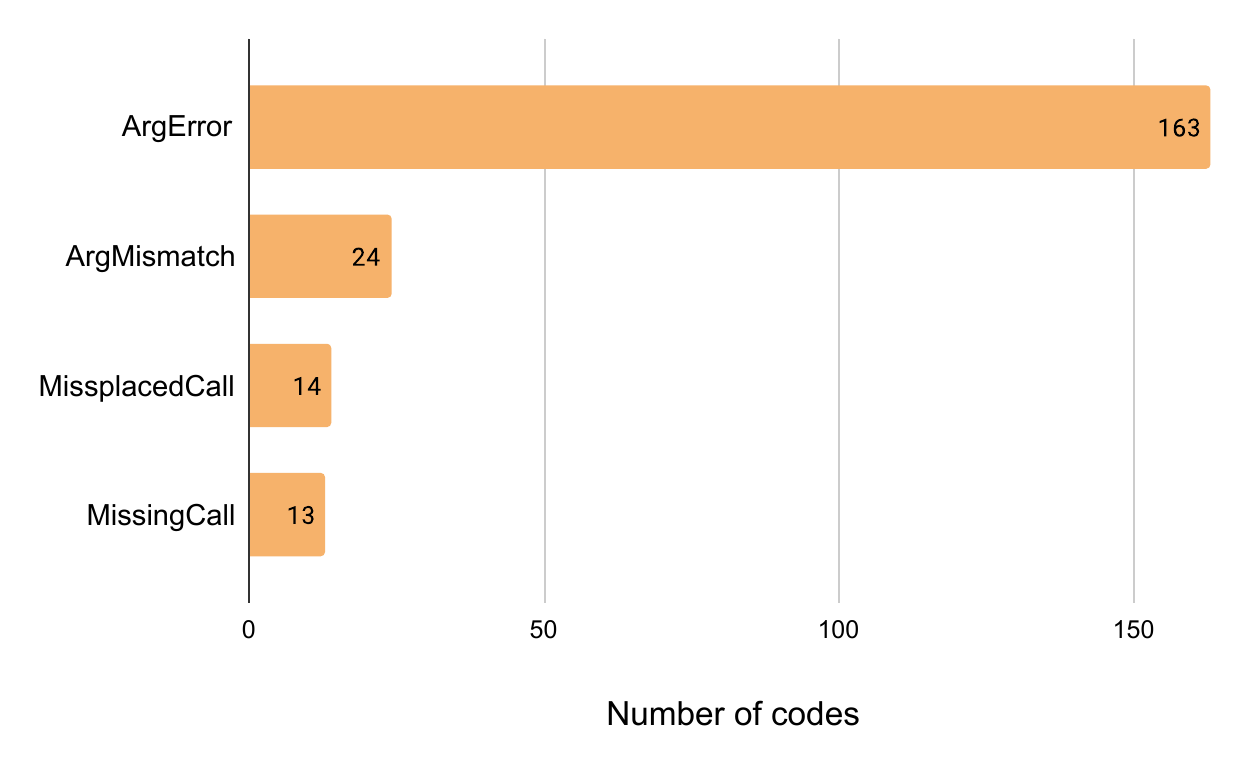}}
  \hfill
  \subfigure[MBI]{\includegraphics[width=0.55\linewidth]{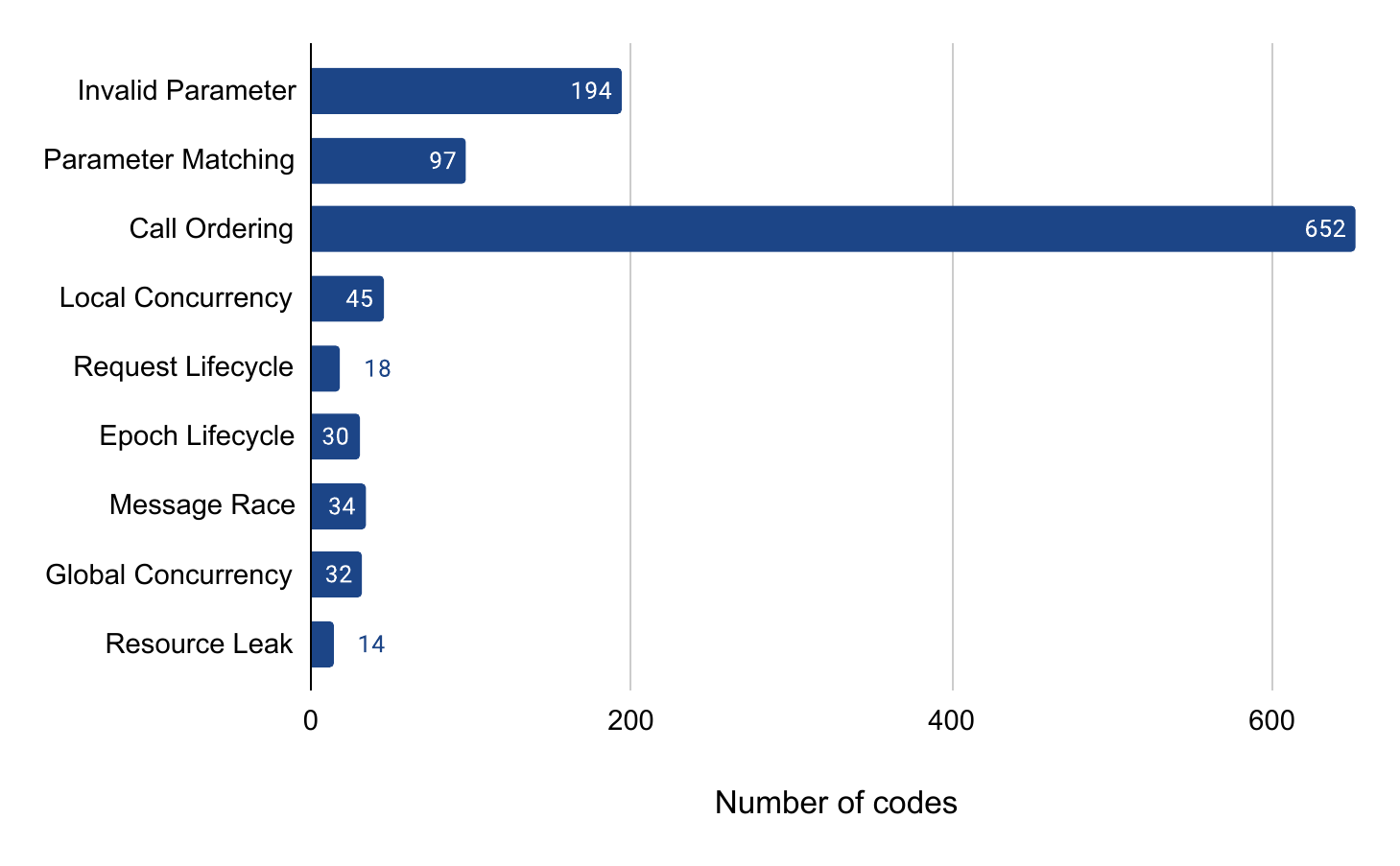}}
  \caption{Number of codes per error type in MPI-CorrBench (left) and MBI (right).}
  \label{fig:statsErrors}
\end{figure*}

\section{Datasets}
\label{sec:dataset}

We created datasets to train ML models by using two MPI benchmarks suites that have recently emerged: the MPI Bugs Initiative (MBI) \cite{mbi} and MPI-CorrBench \cite{corrbench21}. These benchmarks are the only MPI correctness benchmarks and are integrated into state-of-the-art MPI verification tools such as MUST and PARCOACH.

MBI contains almost 2,000 codes written in C. The initiative proposes 9 types of errors, gathered according to which context they manifest: \emph{single call}: invalid parameter, \emph{single process}: resource leak, request lifecycle, epoch lifecycle and local concurrency, and \emph{multi-processes}: parameter matching, message race, call ordering, and global concurrency. Each code in MBI has a header describing the error in the code, how to execute it and what MPI features are used. MPI-CorrBench contains around 400 small codes (referred as level zero) written in C. MPI-CorrBench follows a different error classification compared to MBI. MPI-CorrBench errors can be either erroneous arguments (ArgError), mismatching arguments (ArgMismatch) or erroneous program flow (MissplacedCall and MissingCall). Unlike MBI, codes in MPI-CorrBench do not have a header specifying the errors. Therefore, we relied upon programs names to associate a program with its error type (e.g., the code \texttt{ArgError-MPIIRecv-Count-1.c} is associated with the ArgError error). Both benchmark suites also include a substantial number of correct codes, which is crucial to test our models across a range of different contexts and use cases.



\begin{figure*}[ht!]
  \centering
  \subfigure[MPI-CorrBench]{\includegraphics[width=0.49\linewidth]{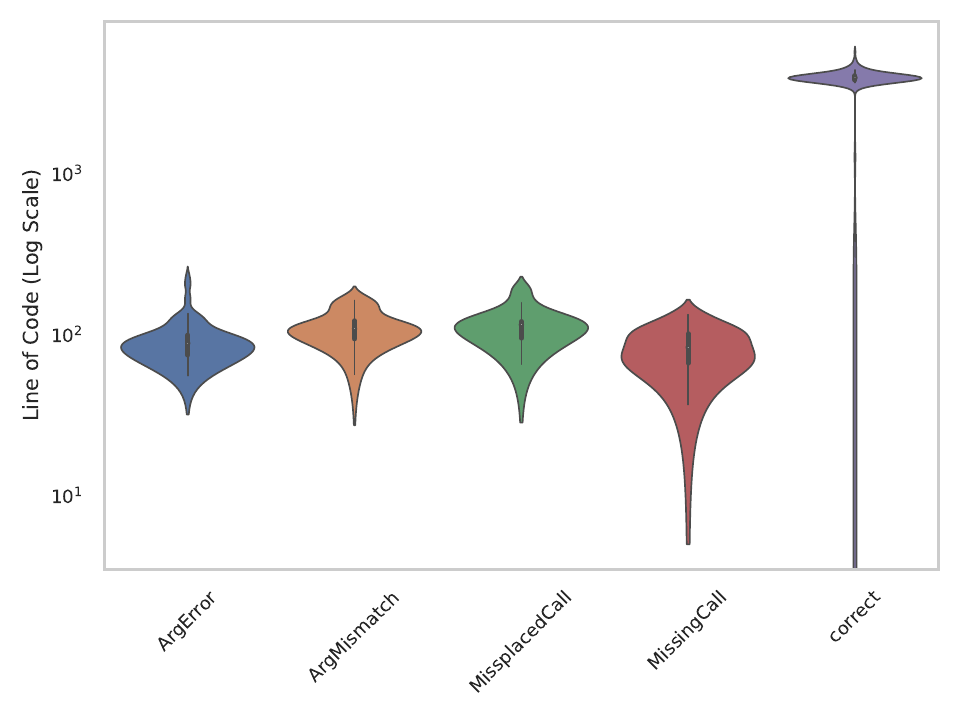}}
  \hfill
  \subfigure[MBI]{\includegraphics[width=0.49\linewidth]{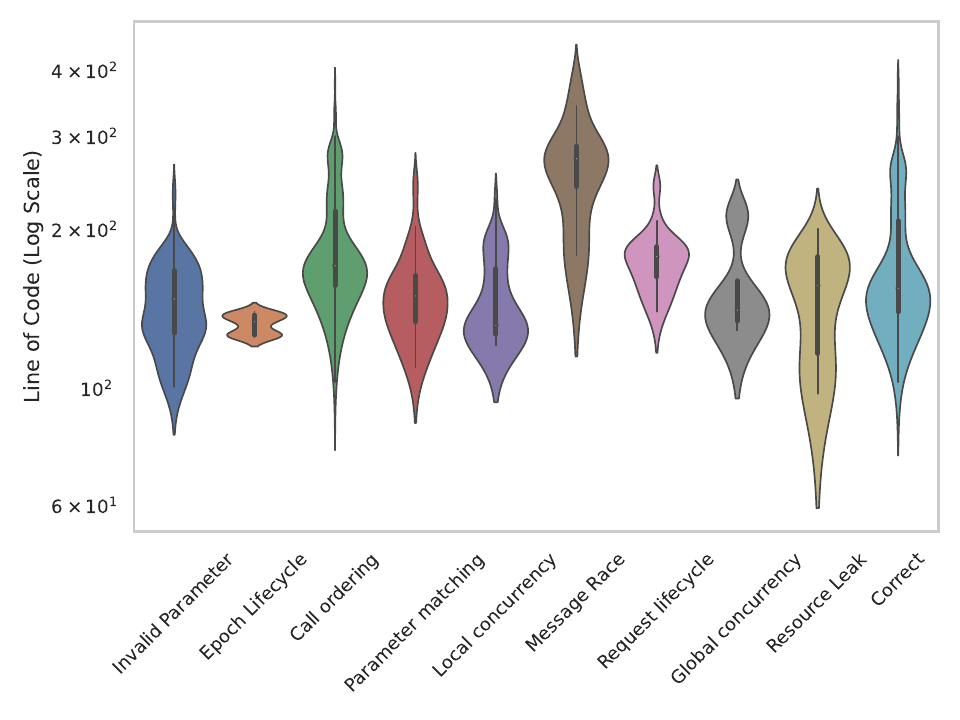}}
  \caption{Code size in MPI-CorrBench (left) and MBI (right). The line of code is reported after performing the C pre-processing include calls. MPI-CorrBench correct codes have a high line count compared to the incorrect codes. On the opposite, MBI has no significant outlier in the line count.}
  \label{fig:statslines}
\end{figure*}

\begin{figure}[h!]
  \centering
\includegraphics[width=0.5\textwidth]{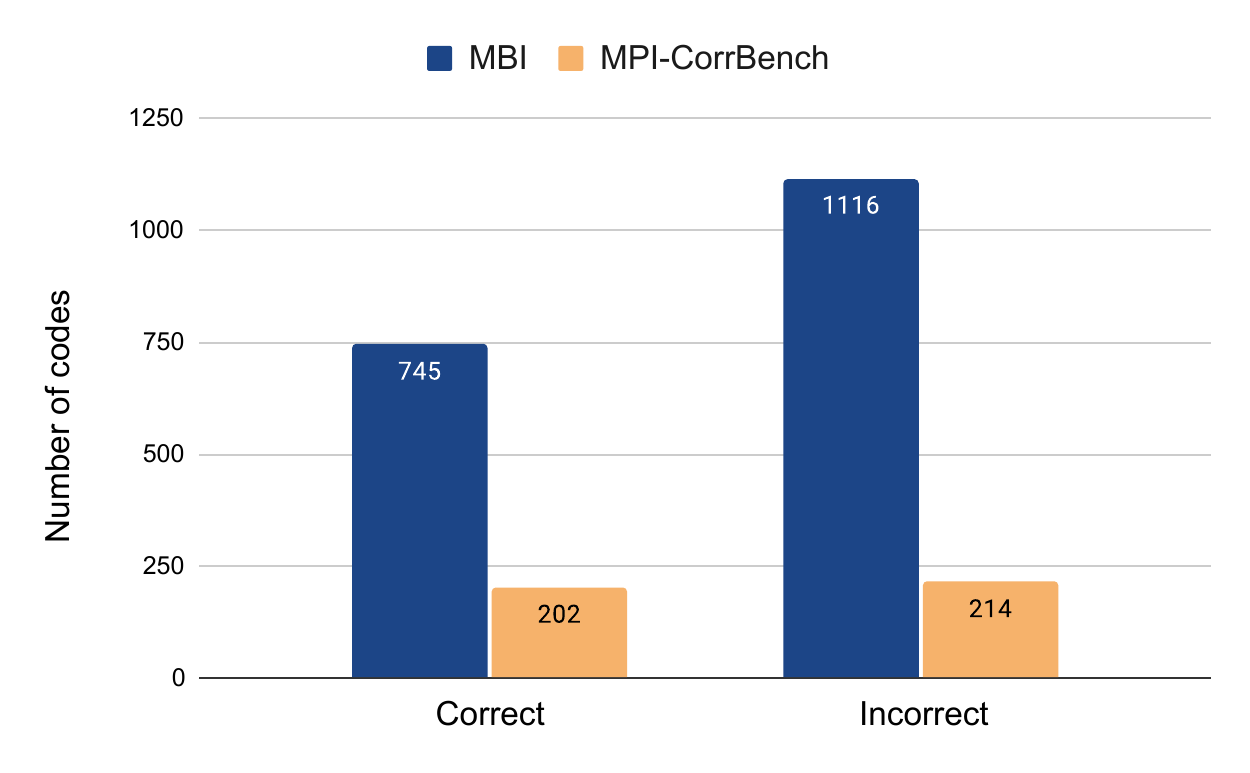}
  \caption{Number of correct and incorrect codes in MBI and MPI-CorrBench.}
  \label{fig:statsCORvsINCOR}
\end{figure}

Figure \ref{fig:statsCORvsINCOR} shows the number of correct and incorrect codes in each benchmark. The distribution among incorrect codes is presented in figure \ref{fig:statsErrors}. In both benchmarks, one type of error is more represented: call ordering for MBI and ArgError for MPI-CorrBench. 
To gain a deeper understanding of the structure and composition of the two benchmarks, we use a violin
plot, presented in figure \ref{fig:statslines}, to visualize the number of lines in each individual code in the benchmarks. In MBI, we observe that the codes are relatively similar in length. However, in MPI-CorrBench, the violin plot shows a wider range of
chord lengths, with some codes consisting of just a few lines and others containing much larger and more
complex structures. In particular, unlike incorrect codes, correct codes have at least $10^3$ lines of code in MPI-CorrBench. This raises concerns that the model might be biased toward longer codes to identify correct codes. Code length is a poor criteria
to predict the presence or absence of errors (or at least not an absolute one). To address this issue, we applied several methods to remove
the bias so that the model can accurately distinguish between correct and incorrect codes. Specifically, correct codes include an extra header \textit{"mpitest.h"} in MPI-CorrBench, which is not necessary to compile the applications and adds all the extra lines during the pre-processing: we simply removed this call across the different codes. We also consider various compilation options and normalization methods that we further describe in Section \ref{subsec:intra}.


In this paper, we consider three different datasets for our models: MBI, MPI-CorrBench without bias due to code size (to which we refer simply as MPI-CorrBench for the rest of the paper), and the combination of MBI with MPI-CorrBench, referred as \emph{Mix} in the rest of the paper. 


\section{Machine Learning Methods}
\label{sec:methods}


This section presents how we train and apply our models over the datasets presented in the previous section. 
To predict if an application is correct, our model needs to be trained over a set of codes that are represented each with a set of features along with an error label. The datasets already label each code with an error type or identify it as correct. Different levels of programs representations such as Abstract Syntax (AST), control flow, or data flow are valid options to extract the features. We select the LLVM Intermediate Representation (IR) as input for our models since it proved to be a successful representation to expose information for Deep Learning (DL) models when optimizing various tasks~\cite{cummins2021programl,venkatakeerthy2020ir2vec}. The rest of this section focuses on how each of our model uses the IR along with the dataset labels to train models. 

\subsection{IR2vec embedding}
\label{sec:ir2vec}

\begin{figure*}[htbp]
\centering
\includegraphics[width=\textwidth]{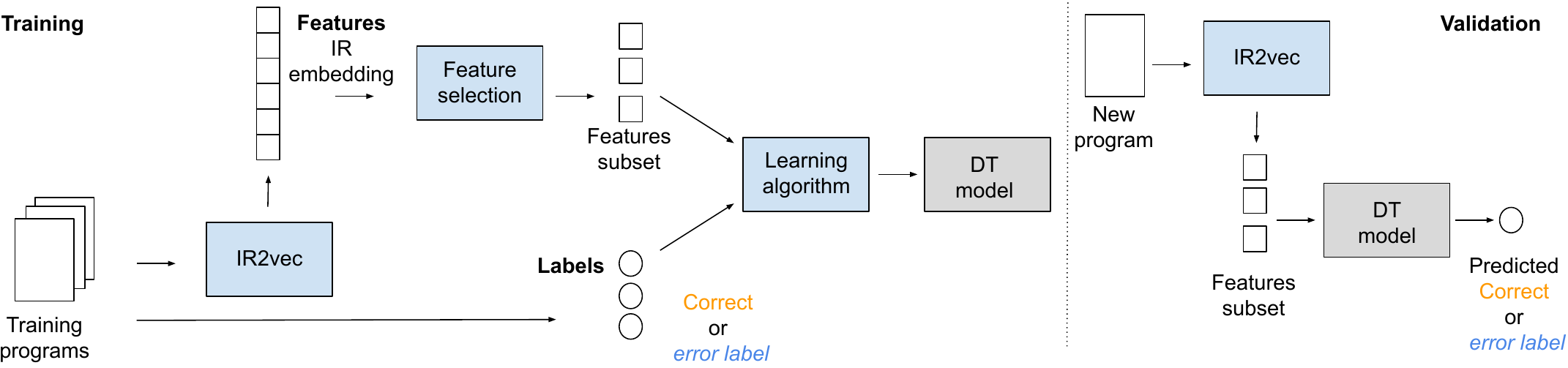}
        \caption{Predicting errors in MPI applications with embedding based models. }
        \label{fig:ir2vec-workflow}
\end{figure*}

Our first strategy consists in exposing the IR to ML models through embedding with IR2vec~\cite{venkatakeerthy2020ir2vec}. 
Figure~\ref{fig:ir2vec-workflow} presents the detailed workflow of how we train and validate the IR2vec based prediction model. IR2vec transforms IR code into a vector embedding in a continuous space. The underlying assumption behind this transformation is that \textit{similar} applications should result in vectors that are close to each other. IR2vec was previously applied for optimizing heterogeneous task mapping and thread coarsening but not for verification: In this work, we extend the IR embedding for error detection. 

IR2vec proposes 2 encoding strategies, symbolic and flow aware. The former performs a seed embedding while the latter performs a seed embedding but also augments it with flow-aware information. Each encoding generates a vector of $256$ elements per IR compilation unit. 

Because the cost of inferring the embedding is negligible compared to executing an MPI application, we executed both the symbolic and flow aware encodings and concatenated them into a single vector. We use this vector as feature input for a ML classifier to determine if a code is correct or contains a specific error.
In particular, we provide the concatenated embedding vector to a Decision Tree (DT). The labels of the decision tree are simply the different types of error described in Section~\ref{sec:dataset} (or a description of whether the code is correct or not) while the features are the vector embeddings for each code. The DT uses the default Scikit learn setup~\cite{pedregosa2011scikit} (version 1.0).

To remove the noise in the feature vectors, we perform a feature selection step with Genetic Algorithms (GAs). Each individual is considered as a subset of vector coordinates. The fitness function of the individual is the quality of the subsequent prediction model. We mutate individuals by changing which vector coordinates are selected. In total,  we consider a population size of $2500$ individuals, with $25$ generations, $90\%$ and $10\%$ crossover and mutation probabilities, respectively, where each individual is composed of $5$ vector coordinates. The GA was implemented with \textit{pyeasyga}~\cite{pyeasyga} (version 0.3.1). Section~\ref{subsec:cross} evaluates the benefits of this step.

\subsection{Graph Neural Network Embedding}

\begin{figure*}[htbp]
\centering
\includegraphics[width=\textwidth]{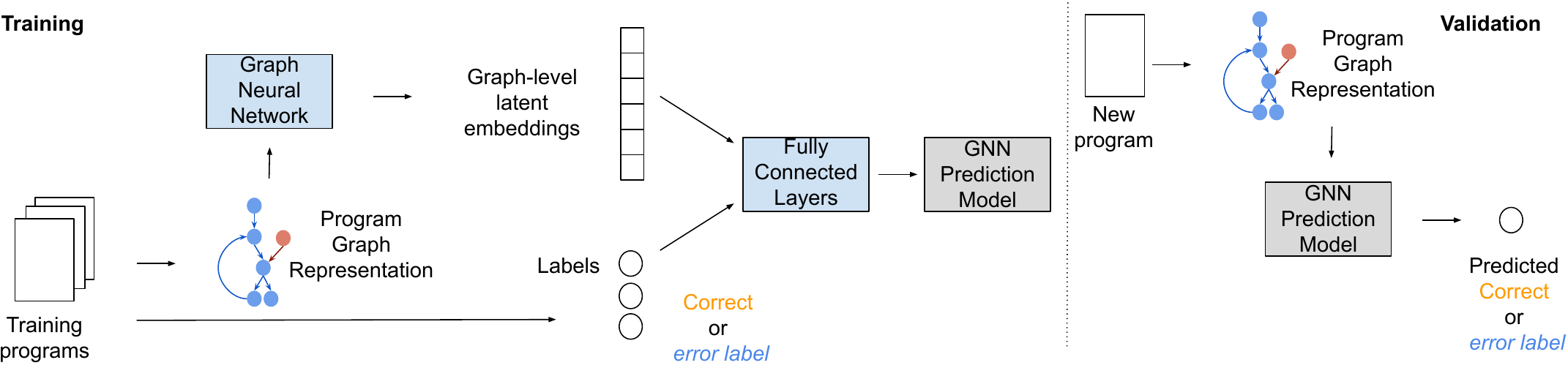}
        \caption{Graph Neural Network based model to predict errors in MPI.}
        \label{fig:graph_correct_overall_workflow}
\end{figure*}

Our second strategy utilizes graphs and Graph Neural Networks (GNNs) to identify MPI bugs as shown in Figure \ref{fig:graph_correct_overall_workflow}.
A lot of software analyses are performed over graphs, such as control flow and data flow.
Therefore, representing programs as graphs allows us to present such flow-aware information to the neural network models. 
To classify programs using GNNs, we adapt ProGraML~\cite{cummins2021programl} representation, which is a program graph representation built on top of LLVM IR. 
It specifically creates data flow, control flow, and call graphs and presents them in one unified graph.
As a result, we have three types of edges or relations between nodes.
To effectively model different relations and nodes, we treat each graph as a heterogeneous graph with three types of nodes (i.e., control, variable, constants) and three types of edges (i.e., control, data, call) and use HeteroConv layers in PyTorch Geometric \cite{fey2019fast} to support heterogeneous graphs. We use Graph Attention Convolution (GATv2) \cite{brody2021attentive} as our graph convolution layer.

Therefore, our GNN-based MPI error detection pipeline is as follows:
We first generate ProGraML representation for each sample in our dataset, then feed these graphs to 3 consecutive GATv2 layers with the sizes of 128, 64, and 32 respectively.
After applying GATv2 layers, we have a latent representation (vector) of size 32 for each node in the graphs. 
Then we apply an adaptive max pooling layer to aggregate the latent representation of all nodes into one vector. As a result, we would have one vector per each graph (graph-level vector).
Then, graph-level vectors are passed to two fully connected layers. The output dimension of the last fully connected layer corresponds to the number of classes that we have in our training set.

We use the cross-entropy loss function to measure the error rate of the GNN model during training and use the Adam optimizer with the learning rate of $4 \times 10^{-4}$ to update the weights of the model to minimize the loss value. The GNN pipeline is trained for 10 epochs.

 \begin{table*}
  \begin{center}
    \begin{tabular}{|c|l|l|}\hline
      \textbf{Metric}&\textbf{Definition} & \textbf{Meaning}\\\hline
      Recall&\multirow{1.4}{*}{$R=\frac{TP}{TP+FN}$}&\multirow{1.4}{*}{Ability to find existing errors}\\[5pt]\hline
      Precision& \multirow{1.4}{*}{$P=\frac{TP}{TP+FP}$} &\multirow{1.4}{*}{Potential confidence when a code is reported as correct}\\[5pt]\hline
      F1 Score&\multirow{1.4}{*}{$F1=\frac{2 \times P \times R}{P+R}$} & \multirow{1.4}{*}{Overall bug-finding quality}\\[5pt]\hline
      Accuracy&\multirow{1.4}{*}{$A=\frac{TP+TN}{Total}$} & \multirow{1.4}{*}{Proportion of correct diagnostics over the tests}\\[5pt]\hline\hline
      Coverage&\multirow{1.4}{*}{$Cov=1-\frac{CE}{Total + Errors}$} & \multirow{1.4}{*}{Ability to compile codes}\\[5pt]\hline
      Conclusiveness&\multirow{1.4}{*}{$Cc=1-\frac{Errors}{Total+Errors}$} & \multirow{1.4}{*}{Ability to draw a diagnostic on codes}\\[5pt]\hline
     Specificity&\multirow{1.4}{*}{$S=1-\frac{TN}{TN+FP}$} & \multirow{1.4}{*}{Ability to not find errors in correct codes}\\[5pt]\hline
      Overall accuracy&\multirow{1.4}{*}{$Oa=\frac{TP+TN}{Total+Errors}$} & \multirow{1.4}{*}{Proportion of correct diagnostics over all tests}\\[5pt]\hline
    \end{tabular}
    \vspace{.1cm}
    \caption{Metrics used in the evaluation. CE = Compilation Error, TO = Time Out, RE = Runtime Error, TP = True Positive, TN = True negative, FP = False Positive, FN = False Negative. Total = TP+FP+TN+FN , Errors = CE+RE+TO. The second part of the table depicts metrics defined in \cite{mbi}.}
    \label{tab:metrics}
  \end{center}
\end{table*}

\begin{table*}[h]
    \centering
\setlength\tabcolsep{2pt} 
\begin{tabular}{|l|*{2}{c|}|*{4}{c|}|*{4}{c|}}\hline
  \multirow{2}{*}{ \textbf{Model}} &  \multicolumn{2}{c||}{Dataset} &\multicolumn{4}{c||}{Results} &\multicolumn{4}{c|}{Metrics}\\\cline{2-11}
& \textbf{Training}&\textbf{Validation}  & \textbf{TP} & \textbf{TN} & \textbf{FP} & \textbf{FN} & \textbf{Recall}& \textbf{Precision}& \textbf{F1 Score}  & \textbf{Accuracy}  \\\hline 
    \multirow{2}{*}{IR2vec Intra }  &  MBI  & MBI & 1043 & 664 & 81  & 73     & 0.935 & 0.928 & 0.931 & 0.917 \\\cline{2-11} 
                                    &  CORR  & CORR & 200  & 184 & 18 & 14    & 0.934 & 0.917 & 0.925 & 0.923\\\hline 
    \multirow{2}{*}{IR2vec Cross}   &  MBI  & CORR & 182  & 176  & 32 &  26  & 0.875 & 0.850 & 0.862 & 0.860 \\\cline{2-11} 
                                    &  CORR  & MBI &  805 & 523 & 311 & 222    & 0.784 & 0.721 & 0.751 & 0.713 \\\hline 
    IR2vec Mix      & MBI + CORR &  MBI + CORR   &  1198 & 811 & 136  & 132   & 0.901 & 0.898 & 0.899 & 0.882 \\\hline\hline 
    \multirow{2}{*}{GNN Intra}      &  MBI  & MBI & 1045 & 657 & 88  &  71    & 0.922 & 0.936 & 0.929 & 0.914 \\\cline{2-11} 
                                    &  CORR  & CORR & 180 	 & 154 & 22  & 60 & 0.719 & 0.875 & 0.79 & 0.803  \\\hline 
    \multirow{2}{*}{GNN Cross}      & MBI  & CORR & 189	 & 168 & 34  & 25   & 0.832 & 0.870 & 0.851  & 0.858  \\\cline{2-11} 
                                    &  CORR  & MBI & 1108 & 19 &  726 & 8   & 0.604 & 0.993 & 0.751  & 0.605  \\\hline 
    GNN Mix                         & MBI + CORR &  MBI + CORR & 1228	 &846 & 101 & 102    & 0.893 & 0.892 & 0.893 & 0.911   \\\hline 
\end{tabular}
\setlength\tabcolsep{6pt} 
\vspace{.1cm}
  \caption{Results of our models on the three datasets. CORR = MPI-CorrBench. All the predictions are on Whether the code is correct or incorrect.}
  \label{tab:synthesizedResults}
\end{table*}

\begin{table*}[ht]
    \centering
\setlength\tabcolsep{2pt} 
\begin{tabular}{|l|*{3}{c|}|*{4}{c|}|*{2}{c|}|*{4}{c|}|c|}\hline
  \multirow{2}{*}{ \textbf{Tool}} &  \multicolumn{3}{c||}{Errors} &\multicolumn{4}{c||}{Results}&\multicolumn{2}{c||}{Robustness} &\multicolumn{4}{c||}{Usefulness}&\textbf{Overall}\\\cline{2-14}
& \textbf{CE}&\textbf{TO}&\textbf{RE}  & \textbf{TP} & \textbf{TN} & \textbf{FP} & \textbf{FN} &\textbf{Coverage} & \textbf{Conclusiveness} & \textbf{Specificity}&\textbf{Recall}& \textbf{Precision}& \textbf{F1 Score}    & \textbf{accuracy}\\\hline 
ITAC            & 0 & 157 & 1 &  859 & 738 & 4   & 102 & 1 & 0.915 & \textbf{0.995} & 0.894 & \textbf{0.995} & \textbf{0.942} & 0.858 \\\hline
PARCOACH V2.3.1 & 0 &  0  & 0 &  775 & 66  & 679 & 341 & 1 &   1   & 0.088 & 0.694 & 0.533 & 0.603 & 0.452
\\\hline
\hline
IR2vec Intra    & 0 &  0  & 0 & 1043 & 664 & 81  & 73  & 1 &   1   & 0.891 & \textbf{0.935} & 0.928 & 0.931 & \textbf{0.917} \\\hline
IR2vec Cross    & 0 &  0  & 0 &  805 & 523 & 311 & 222 & 1 &   1   & 0.627 & 0.784 & 0.721 & 0.751 & 0.714 \\\hline
GNN Intra       & 0 &  0  & 0 & 1045 	 & 657 & 88  & 71  & 1 &   1   & 0.902 & 0.922 & 0.936 & 0.929 & 0.853 \\\hline
GNN Cross       & 0 &  0  & 0 & 1108	 & 19 & 726  & 8  & 1 &   1   & 0.703 & 0.604 & 0,993  & 0,751 & 0,830 \\\hline
\hline
\textit{Ideal tool}&\textit{0}&\textit{0}&\textit{0}&\textit{1116}&\textit{745}&\textit{0}&\textit{0}&\textit{1}&\textit{1}&\textit{1}&\textit{1}&\textit{1}&\textit{1}&\textit{1} \\\hline
\end{tabular}
\setlength\tabcolsep{6pt} 
\vspace{.1cm}
  \caption{Detailed Methods Evaluation against the \textsc{MPI Bugs Initiative}. Best results are in bold. All the predictions are on Whether the code is correct or incorrect.}
  \label{tab:MBIResults}
\end{table*}

\section{Experimental Results}
\label{sec:results}

This section presents the results of our two models on the datasets, a comparison with related works, and an ablation study for generalization purposes. We used MBI (commit \textit{1ab5a546}) and MPI-CorrBench version 1.2.1. Our results are reproducible at   \url{https://gitlab.inria.fr/reproducibility/paper-mpi-errors-detection-using-gnn-embedding-and-vector-embedding-over-llvm-ir-reproducibility}.

We designed different evaluation scenarios to incrementally increase the difficulty of detecting errors: \textit{Intra}, \textit{Mix} (previously described in Section~\ref{sec:dataset}), and \textit{Cross}. \textbf{Intra} (see Section~\ref{subsec:intra}) consists in training and evaluating models on either MBI or MPI-CorrBench while \textbf{Mix} (see Section~\ref{subsec:mix}) consider both benchmark suites together. Such scenarios are helpful to determine if ML based approaches are actually capable of detecting errors even on applications that come from the same benchmark suite and thus share some code structures. To avoid over-fitting, we used a standard \textit{10-fold cross-validation} to evaluate our models over Intra and Mix. 
Thus, for each dataset, ML method (e.g., IR2vec and GNN), and Intra and Mix, we train ten models and evaluate each one over validation folds composed of approximately 200, 40, and 240 codes (i.e., 10\%) for MBI, MPI-CorrBench, and mix, respectively. For the rest of the paper, all the prediction results over Intra or Mix are an aggregation of the $10$ validation folds.

Finally, \textbf{Cross} (see Section~\ref{subsec:cross}) refers to train a model on a benchmark suite and evaluate it on a distinct benchmark suite (e.g., train model on MBI and validate it on MPI-CorrBench). This scenario demonstrates the generalization capability of our models by evaluating them on significantly different code structures and errors: not only are the code structures changing between benchmark suites, but so are the labelled errors (as discussed in Section~\ref{sec:dataset}). This is particularly challenging for the models as the different datasets might literally not contain some errors, limiting the potential accuracy of our approach. To train and evaluate the same model over different datasets, we updated the labeling of each code to either correct or incorrect. 

To assess the quality of our predictions when considering correct or incorrect labels, we used a set of metrics that Table~\ref{tab:metrics} summarizes. Each metric is computed with the number of true positive \textit{TP} (error correctly detected), true negative \textit{TN} (correct code reported as such), false positive \textit{FP} (correct code reported as faulty) and false negative \textit{FN} (error missed). The first part of the table gives the most used metrics in the state-of-the-art while the second part of the table gives metrics defined in MBI.

\subsection{Intra Modeling}
\label{subsec:intra}

We first trained and validated our models on a standalone benchmark suite (MBI or MPI-CorrBench). 
End results of correct and incorrect code predictions are presented in Table~\ref{tab:synthesizedResults} under the lines \emph{IR2vec Intra} and \emph{GNN Intra}. To achieve these results, we explored different parameters in the models. In particuar, we consider compiler options for both approaches, and normalization, prediction labels, feature selection, and seeds for IR2vec. These are parameters in our methods that can be fine-tuned to increase the overall accuracy or provide more insights. 

\textbf{Compilation options.} Both GNN and IR2vec methods use IR as input. It is interesting to note that the compiler can optimize the IR with different compiler passes (e.g., \textit{-O2}, \textit{-O3}) to improve performance. However, recent studies~\cite{TehraniJamsaz2022ipdps,li2022unleashing} demonstrated that compiler passes also enable to expose more information to DNNs. DNNs~\cite{TehraniJamsaz2022ipdps,li2022unleashing} take as input IR and use it for parameter optimization such as device mapping or NUMA setting. In other words, using custom compiler passes with DNNs improve the prediction capabilities over a single compiler sequence when generating the IR. In our work, we considered different compiler options and selected two in particular for the rest of this work: \textit{-O0} for GNN and \textit{-OS} for IR2vec. Our intuition is that since \textit{-O0} leaves the code intact, it may ease the error detection. However, some errors may only occur when the code is optimized (i.e., a bug might not be visible, but optimizations may trigger its manifestation): there is a trade-off between studying a simplified code versus one complex that is representative of the real execution. We selected \textit{-OS} for IR2vec to reduce the impact of different code sizes: our assumption is that \textit{-OS} will reduce the IR size difference between different codes. We empirically evaluated \textit{-O0} (easy to analyze), \textit{-O2} (representative), and \textit{-OS} (reduced bias due to size) in Table~\ref{detail-intra}. At most, compiler optimizations improve the accuracy by approximately $5\%$ over MPI-CorrBench.

\begin{table*}[h]
    \centering
\setlength\tabcolsep{2pt} 
\begin{tabular}{|c|c|c|*{4}{c|}|*{4}{c|}}\hline
  \multirow{2}{*}{\textbf{Compilation option}} &  \multirow{2}{*}{\textbf{Normalization}} & \multirow{2}{*}{\textbf{Dataset}} &\multicolumn{4}{c||}{Results} &\multicolumn{4}{c|}{Metrics}\\\cline{4-11}
  & & & \textbf{TP} & \textbf{TN} & \textbf{FP} & \textbf{FN} & \textbf{Recall}& \textbf{Precision}& \textbf{F1 Score}  & \textbf{Accuracy}  \\\hline 
  -O0 & \multirow{3}{*}{none}    & \multirow{3}{*}{MBI} & 1042&	681&	64&74     & 0.934	&0.942	& 0.938&	0.926 \\ 
-O2 &     &  &1039&	662&	83&	77    &  0.931	&0.926	&0.929&	0.914 \\ 
-Os &     &  & 1047	&665&	80	&69    & 0.938 &	0.929&	0.934&	0.920 \\\cline{1-2}\cline{3-11} 
-O0 & \multirow{3}{*}{none}   & \multirow{3}{*}{CORR} & 205	&191	&11 & 9    & 0.9579& 0.9491	&0.9535	&0.9519  \\ 
-O2&     &  & 198&	180	&22	&16     & 0.9252&	0.9000&	0.9124	&0.9087  \\ 
-Os&    &  &203&	189&	13&	11     & 0.9486	&0.9398&	0.9442&	0.9423 \\\hline 
-O0& \multirow{3}{*}{vector}     & \multirow{3}{*}{MBI} & 1038	&679&	66	&78    & 0.930	&0.940	&0.935&	0.923 \\ 
-O2&     &  & 1029&	659	&86&	87   & 0.922&	0.923&	0.922&	0.907 \\ 
-Os&     &  &1043	&664&	81&	73    & 0.935&	0.928&	0.931&	0.917\\\cline{1-2}\cline{3-11} 
-O0& \multirow{3}{*}{vector}     & \multirow{3}{*}{CORR} & 202&	186&	16&	12  & 0.9439 &0.9266&0.9352& 0.9327 \\
-O2&     & & 194	&187&	15&	20   &  0.9065	&0.9282&	0.9173&	0.9159  \\
-Os&     & & 200	&184	&18&	14  & 0.9346&	0.9174&	0.9259&	0.9231\\\cline{1-2}\cline{3-11}
-O0& \multirow{3}{*}{index}     & \multirow{3}{*}{MBI} & 1042&	681&	64&	74   & 0.934&	0.942&	0.938	&0.926 \\ 
-O2&    &  &1039&	662&	83&	77    & 0.931&	0.926&	0.929&	0.914  \\ 
-Os&     &  &1047	&665&	80	&69    &0.938&	0.929&	0.934	&0.920 \\\hline 
-O0& \multirow{3}{*}{index}     & \multirow{3}{*}{CORR} & 205&	188	&14	&9    & 0.9579&	0.9361&	0.9469&	0.9447 \\
-O2 &    &  & 202&	188&	14&	12   & 0.9439&	0.9352&	0.9395&	0.9375\\ 
-Os&     &  & 203&	189&	13&	11    & 0.9486&	0.9398&	0.9442&	0.9423\\\hline 
\end{tabular}
\setlength\tabcolsep{6pt} 
\vspace{.1cm}
  \caption{Results for IR2vec Intra with different compilation and normalization options. CORR refers to MPI-CorrBench. All the predictions are on Whether the code is correct or incorrect.}
  \label{detail-intra}
\end{table*}

\textbf{Normalization.} IR2vec generates vectors as input to a DT that actually predicts the code correctness. Yet, naively using this vector can bias the prediction. For instance, we observe that long codes tend to generate larger vectors. This was particularly a problem when studying MPI-CorrBench correct codes: vectors with large values were systematically referring to correct codes due to the bias in the dataset before removing the headers. To further remove such artifacts, we considered two normalization strategies. We either normalized each vector to contain values between $0$ and $1$ by dividing each element with the max of the vector or we normalized each vector coordinate across all the codes. We ended up using the former normalization as it ensures that every code has a vector with values between $0$ and $1$ independently of its size or any other code. Table~\ref{detail-intra} presents in details the different normalization strategies over MBI and MPI-CorrBench. In particular, \textit{none}, \textit{vector}, and \textit{index} refer to no normalization, normalization across the vector, or per element. The impact of optimizing normalization is less than $3\%$ across all the scenarios.

\textbf{Feature selection.} As described in Section~\ref{sec:ir2vec}, features have a significant impact on the prediction accuracy. In addition to normalizing the features, we applied a feature selection process where we only select a subset of the features to remove the noise with GA. Table~\ref{detail-GA} presents the benefit of running a feature selection with GA over naively using the vector. Results were obtained with the $-Os$ and $vector$ as compilation and normalisation options, respectively. Interestingly, feature selection improves the accuracy by $5\%$ and up to $47\%$ for Intra and Cross respectively. This means that the Cross scenario is more sensitive to this parameter.

\begin{table*}[h]
    \centering
\setlength\tabcolsep{2pt} 
\begin{tabular}{|l||c|*{2}{c|}|*{4}{c|}|*{4}{c|}}\hline
  \multirow{2}{*}{ \textbf{Model}} & \multirow{2}{*}{ \textbf{GA}} & \multicolumn{2}{c||}{Dataset} &\multicolumn{4}{c||}{Results} &\multicolumn{4}{c|}{Metrics}\\\cline{3-12}
& & \textbf{Training}&\textbf{Validation}  & \textbf{TP} & \textbf{TN} & \textbf{FP} & \textbf{FN} & \textbf{Recall}& \textbf{Precision}& \textbf{F1 Score}  & \textbf{Accuracy}  \\\hline 
    \multirow{4}{*}{IR2vec Intra }  & OFF &  \multirow{2}{*}{MBI}  & \multirow{2}{*}{MBI} & 989	&635&	110&	127 & 0.886&	0.900&	0.893	& 0.873\\\cline{5-12} 
   & ON & &  &  1043&	664&	81&	73  &  0.935&	0.928&	0.931&	0.917\\\cline{2-12} 
   & OFF &   \multirow{2}{*}{CORR} & \multirow{2}{*}{CORR} & 190&	174	&28	&24      & 0.888 &	0.871 &	0.879&	0.875 \\\cline{5-12}
   & ON &    &  & 200&	184&	18&	14       & 0.935&	0.917&	0.926&	0.923 \\\hline\hline
    \multirow{4}{*}{IR2vec Cross}   & OFF &  \multirow{2}{*}{MBI}  & \multirow{2}{*}{CORR} & 151	&92	&63	&110     & 0.578&	0.706&	0.636&	0.584 \\\cline{5-12}
    & ON &    &  &  182&	176	&32	&26     & 0.875	&0.850&	0.863&	0.861 \\\cline{2-12}
    & OFF &  \multirow{2}{*}{CORR}  & \multirow{2}{*}{MBI} & 955&	235	&161&	510      &  0.652&	0.856&	0.740&	0.639 \\\cline{5-12}
    & ON &    &  & 805&	523&	311&	222     &  0.784&	0.721&	0.751&	0.714 \\\hline
\end{tabular}
\setlength\tabcolsep{6pt} 
\vspace{.1cm}
  \caption{Results for IR2vec Intra and Cross with and without GA. CORR refers to MPI-CorrBench. All the predictions are on Whether the code is correct or incorrect.}
  \label{detail-GA}
\end{table*}

\textbf{Seeds.} We also noticed that the seed used in our model to generate the embedding in IR2vec can have a significant impact on the subsequent predictions. In particular, we performed a GA exploration to select relevant features over an original seed and then we re-generated vectors with IR2vec using a different seed. With \textit{-Os} compilation and \textit{vector} normalization, changing the seed of the embedding resulted in $0.6\%$ losses and no losses for Intra MBI and MPI-CorrBench respectively.

Nevertheless, accuracy losses are expected since the GA has been trained for the original seed. This is particularly the case for Cross where the GA played a critical in the good predictions: we observed accuracy losses of $40.81\%$ and $2.79\%$ when validating over MPI-CorrBench and MBI respectively, indicating that the GA exploration must be adjusted to the embedding in some scenarios (i.e., predicting from MBI to MPI-CorrBench). 




\begin{figure*}[ht!]
  \centering
\includegraphics[width=0.75\linewidth]{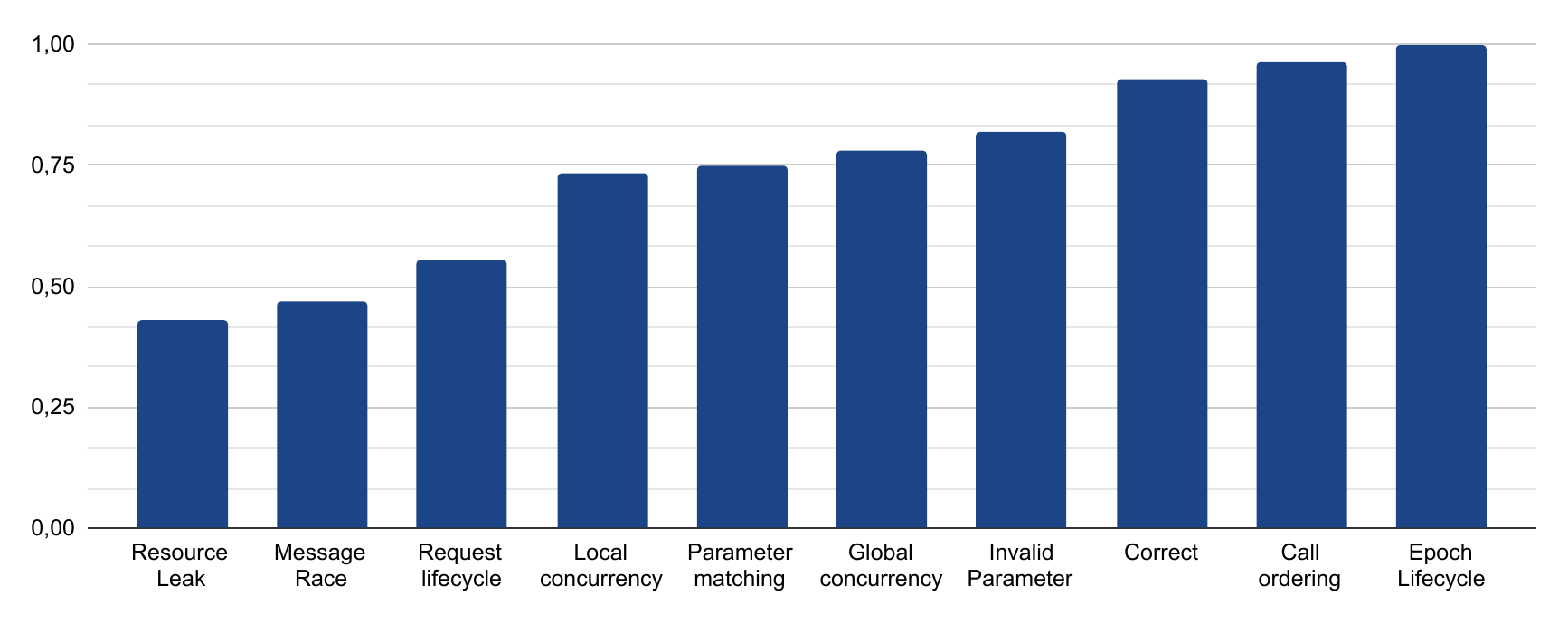}
  \caption{Prediction accuracy of IR2vec per label. We trained a DT classifier with MBI to predict each label separately. Accuracy is calculated as the number of labels correctly predicted in the validation folds divided by the total number of codes with that label. The label prediction quality significantly depends on error types. 
  \label{detail-error}}
\end{figure*}

\begin{figure*}[ht!]
  \centering
  \subfigure[MPI-CorrBench]{\includegraphics[width=0.49\linewidth]{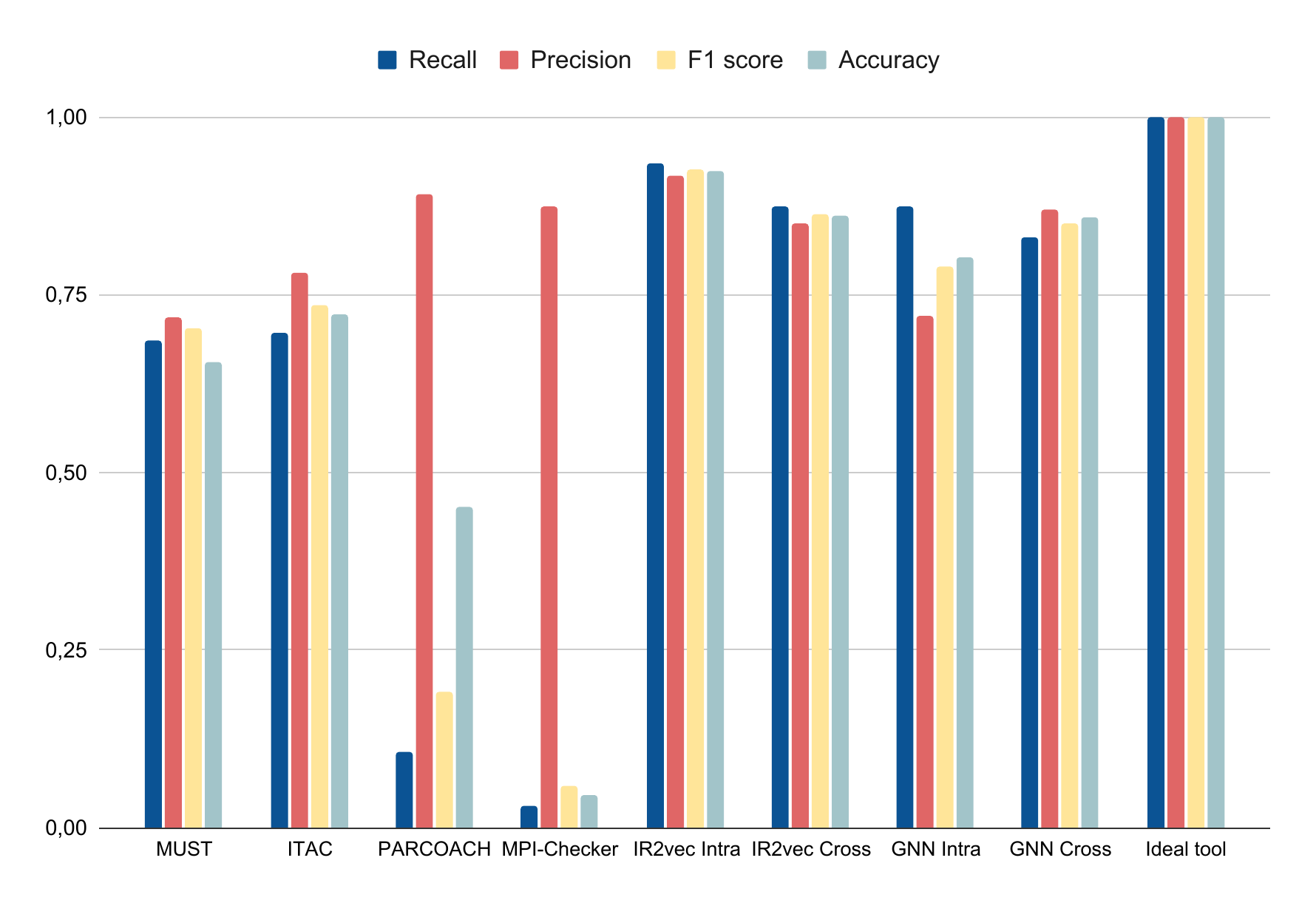}}
  \subfigure[MBI]{\includegraphics[width=0.5\linewidth]{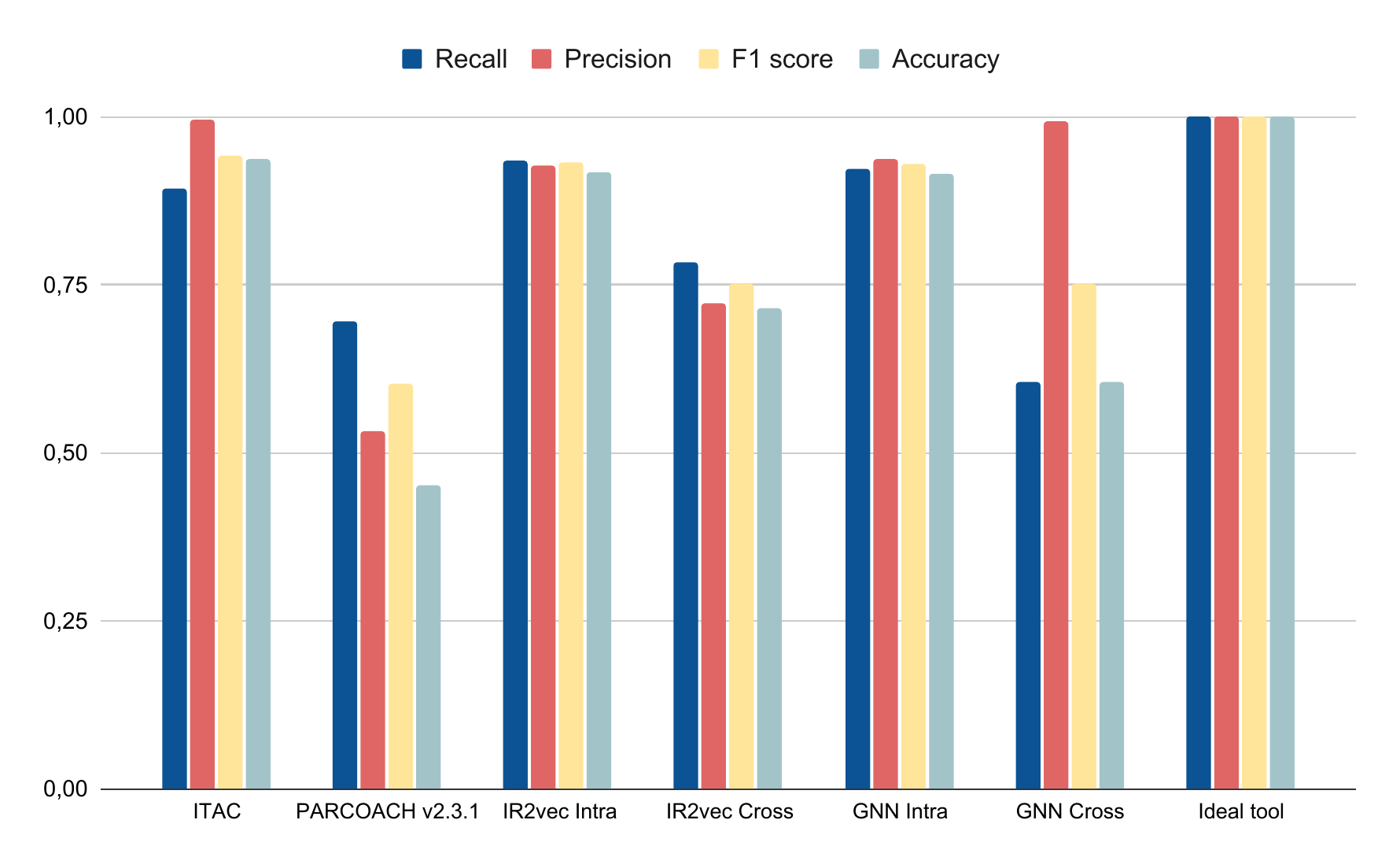}}
  \caption{Metrics Results on MPI-CorrBench (left) and MBI (right). For MPI-CorrBench, results of MUST, ITAC, PARCOACH and MPI-Checker are coming from \cite{corrbench21}.}
  \label{fig:results}
\end{figure*}

\begin{figure*}[ht!]
  \centering
  \subfigure[MPI-CorrBench]{\includegraphics[width=0.49\linewidth]{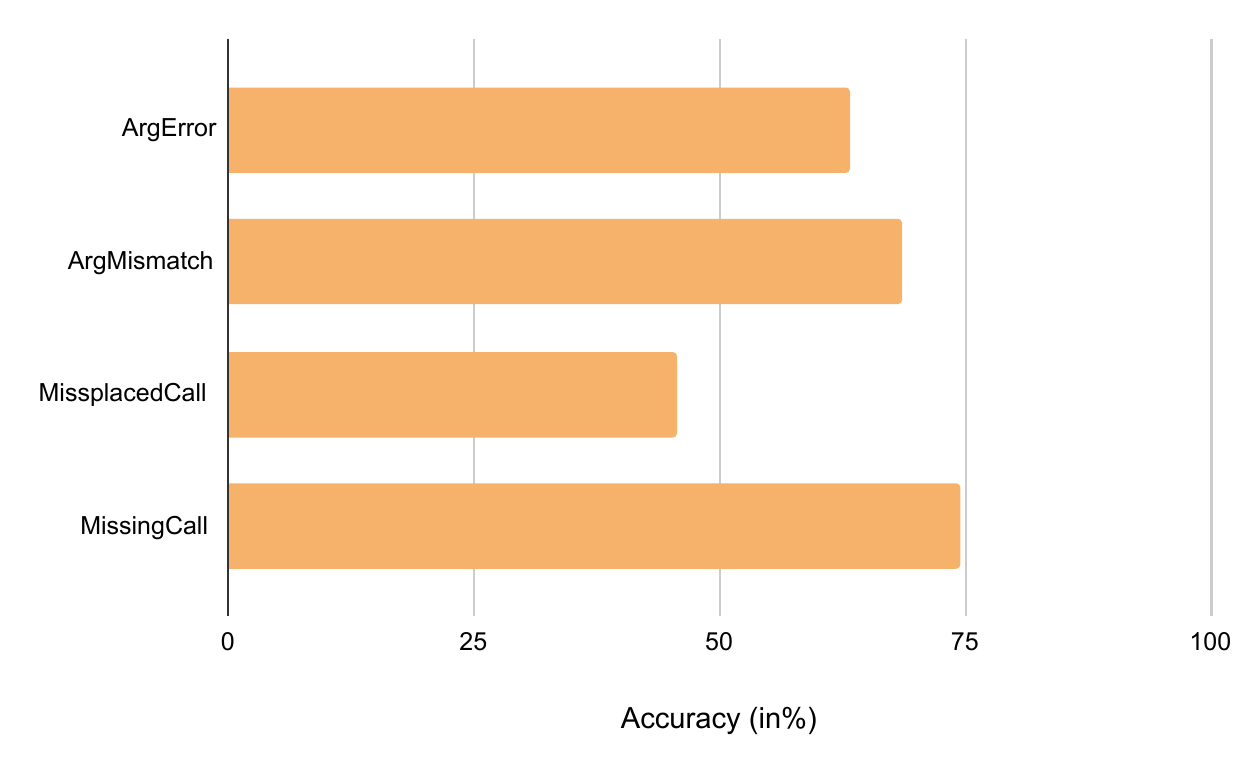}}
  \subfigure[MBI]{\includegraphics[width=0.5\linewidth]{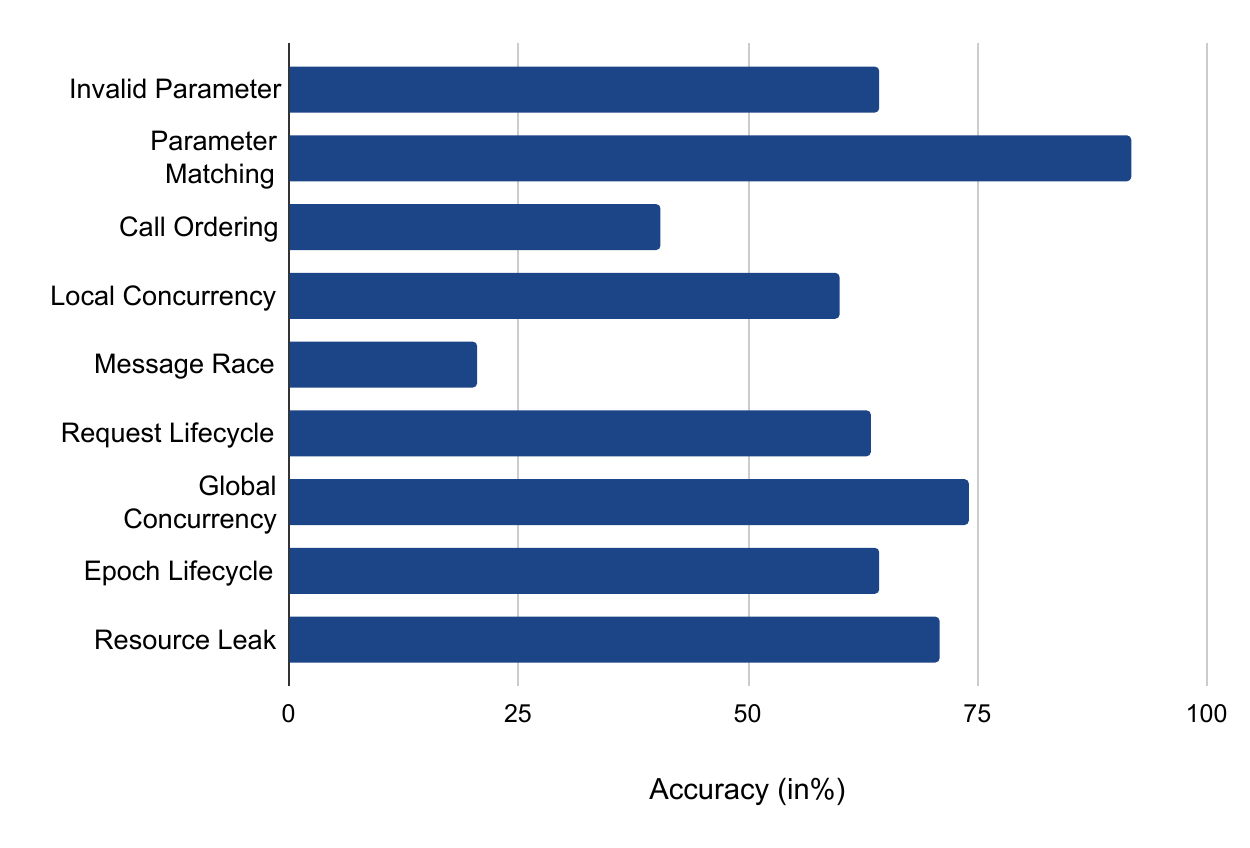}}
  \caption{Ablation study results for MPI-CorrBench (left) and MBI (right).}
  \label{fig:EXresults}
\end{figure*}

\textbf{Prediction labels.} So far, we trained models to determine if a code is correct or incorrect. However, we can further expand our predictions by determining the actual error type instead of just if the code is correct or incorrect. In particular, we trained our DT using the labels of the datasets to directly predict the error type. Please note that this approach is not possible for Cross as the error types are different between training and validation in that scenario. Figure~\ref{detail-error} presents the prediction accuracy per label of our models over MBI. 

We observe $3$ large categories of prediction errors: accurately predicted labels with accuracy over $90\%$ (e.g., Correct, Call Ordering, and Epoch Lifecycle), labels that are mostly correctly predicted (e.g., Invalid Parameter, Parameter Matching) with accuracies around $75\%$, and completely miss-predicted labels such as Message Race or Resource Leak. To understand why our model miss predict some labels, we investigated the number of occurrences per code label in the dataset. Resource Leak has only $14$ instances in our dataset, making it very difficult to learn for the model across the $10$ verification folds. Interestingly, Message Race has more instances than Epoch Lifecycle (which is perfectly predicted): thus the number of samples is not the only reason for the miss-prediction. We suppose that errors exhibit different code patterns, and some are easier to identify with ML approaches. We further investigate the interaction between the errors in Section~\ref{sec:abl}. 

\textbf{End results.} As shown in Table~\ref{tab:synthesizedResults}, IR2vec outperforms GNN with a recall of $0.935$, a precision of $0.928$ and a F1 score of $0.931$. 
Both models show better results on MBI compared to MPI-CorrBench. This may be explained by the higher number of codes in MBI: the models have more codes in the training set.   

\subsection{Mix Modeling}
\label{subsec:mix}

We further extended our models to operate over the dataset Mix composed of both MBI and MPI-CorrBench. The goal of this scenario is to demonstrate if models can predict errors in a larger and more diverse context. For consistency, we again employed a 10-fold cross-validation. The outcomes from mix training and validation is also presented in Table~\ref{tab:synthesizedResults} (lines \emph{IR2vec Mix} and \emph{GNN Mix}) and mirrors the results achieved in the Intra dataset tests. Specifically, we achieved promising results across various metrics: a recall and F1 score of $0.893$ and an accuracy of $0.911$ for the GNN model. These numbers, being in close alignment with the GNN Intra results, highlight the robustness and reliability of our model. Interestingly IR2vec slightly decreases to $0.882$. It is possible that the GNN method can easily scale to larger datasets. 

\subsection{Cross Modeling}
\label{subsec:cross}

Finally, we trained and validated our methods over distinct datasets. This experiment tests the model's ability to generalize and detect errors in untrained, unseen data. This is important because it is desirable that a model does not only rely on the specific error types it was trained on, but to also be able to detect new error types based on its understanding of the underlying code patterns. Results are referred as \emph{IR2vec Cross} and \emph{GNN Cross} in table \ref{tab:synthesizedResults}. 

GNN models struggled to generalize their learnings across different datasets. This is particularly notable when using MPI-CorrBench as a training dataset and MBI as a validation dataset. We obtained an accuracy score of $0.605$ with the GNN model.
The insights and patterns that the model learned from MPI-CorrBench do not seamlessly translate to MBI. 
Such observation raises pertinent questions about the transferability of knowledge in models, emphasizing the need for further refinement to enhance their cross-benchmark applicability. 

Nevertheless, we note that IR2vec achieved an accuracy of $0.713$ and $0.86$ for MPI-CorrBench and MBI validation respectively. These numbers are promising for such difficult scenarios as both code structures and errors labels are different across the datasets. The GA feature selection was critical in achieving them. Indeed, while feature selection moderately impacts Intra predictions, it has a significant impact when cross predicting. In particular, we observed that feature selection improved the accuracy by $47\%$ and $12\%$ when predicting  MPI-CorrBench and MBI respectively.



\subsection{Comparison with Related Works}

Figure~\ref{fig:results} shows the Recall, Precision, F1 score, and Accuracy results of several state of the art verification tools on MPI-CorrBench and MBI. On both subfigures, the last bars depict the results of an ideal tool. 

We investigate MPI-CorrBench in Figure~\ref{fig:results} (a). We used the results presented in~\cite{corrbench21} for MUST, ITAC, PARCOACH and MPI-Checker. 
We compared these results against the prediction of our different models (i.e., IR2vec and GNN) in different scenarios (i.e., Intra and Cross). Our methods outperform the existing verification tools or at least achieve similar results in the most restrictive scenario cross. Our methods achieve a score of at least $0.75$. Furthermore, IR2vec Intra has the closest results to an ideal tool. 

Figure~\ref{fig:results} (b) presents the results over MBI. Because we used a different version of MBI than in~\cite{mbi}, we had to reproduce the experiments for PARCOACH and ITAC. We also used the last versions of these tools. We chose to compare our method only with ITAC and PARCOACH as ITAC is the best tool in \cite{mbi} and PARCOACH is the only static tool used in MBI. \textit{This makes a fair comparison as our approach is also static} (as we also only study the LLVM IR). We observe that ITAC has the best precision, F1 score and accuracy. Yet, IR2vec Intra shows competitive results to ITAC, and more importantly, does not require executing the applications. Indeed, static analyses enable an early detection of errors and avoid the cost of program execution. Therefore, our method can easily be integrated into an automatic toolchain where, at compilation, a light ML-based verification step checks the code.

Table~\ref{tab:MBIResults} further details MBI results. It gives the number of compilation errors (CE), time out (TO), runtime errors (RE), TP, TN, FP and FN as well as seven metrics, defined in \cite{mbi}, depicting the robustness, the usefulness and overall accuracy of the different methods. The last row displays the results of an ideal tool and best results are in bold. The tools and our models have all a coverage of 1 as none of them have compilation error. However, ITAC has 157 time out and 1 runtime error which leads to a conclusiveness (i.e., ability to draw a diagnostic on codes) of $0.915$. ITAC has the best specificity, precision and F1 score whereas IR2vec Intra has the best recall and overall accuracy (and unlike ITAC, operates statically). We further compare the different approaches in Section~\ref{sec:lim}.

\subsection{Ablation Study} 
\label{sec:abl}

The goal of this subsection is to study the interaction between the different error labels. The idea is to remove one labelled error from all the training sets and evaluate if the resulting models can detect it in the validation sets. To do so, we reproduced the $10$ folds cross validation but in addition ensured that no samples of the target label are ever present in the training codes. The model capabilities to predict erroneous codes that it has never seen before demonstrate 1) its generalization to new scenarios as well as 2) the code patterns that are shared between the different error labels. We implemented this study with IR2vec over the datasets presented in Section~\ref{subsec:intra}.

\textbf{Model generalization.} Figure \ref{fig:EXresults} presents results of the ablation study for both MPI-CorrBench and MBI. Each bar represents the prediction accuracy per label (i.e., when the label is excluded from training and only occurs at validation). For each label, we calculate the accuracy as the number of samples of the label correctly predicted as incorrect divided by the total count of that label. Note that we trained the model to predict correct or incorrect and not the label itself as it does not appear in the training codes. 

Labels such as Parameter Matching, MissingCall, or Global Concurrency have a high accuracy score (around or over $75\%$). On the opposite some labels such as Message Race or MissplacedCall are very difficult to generalize over. It is interesting to note that Resource Leak is better predicted in the ablation study than when we directly train models to identify it (see figure~\ref{detail-error} for prediction per label). The ablation model determines if a code is correct or incorrect while the Intra model explicitly predicts it: it is therefore likely that our Intra model confused it with some other related error. 
Overall, our model has the potential to be applied on new errors that it has not encountered before. 

\begin{figure}[ht!]
  \centering
\includegraphics[width=\linewidth]{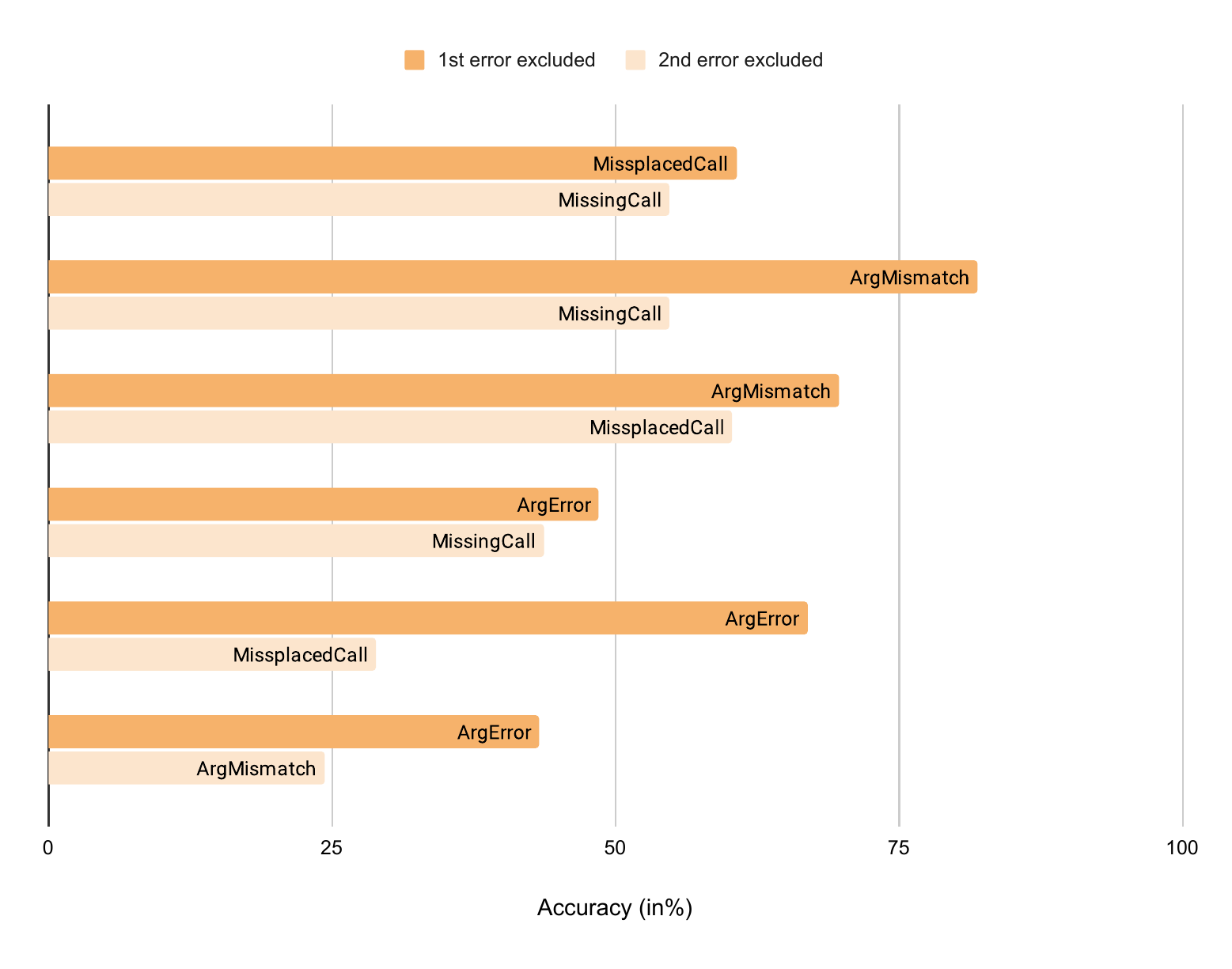}
  \caption{Ablation study results for MPI-CorrBench when two labels are excluded from training. 
  \label{ablation-2labels}}
\end{figure}

\textbf{Error interaction.} Figure \ref{ablation-2labels} shows the prediction accuracy when two labels are excluded from training with MPI-CorrBench as a dataset. While MissingCall was well predicted when excluded from the training set, its accuracy score falls down to $44\%$ when ArgError is also excluded from training. This shows a similarity between the two error types (they exhibit differences at source but similar embeddings help detect them). Conversely, MissplacedCall has a higher accuracy score if ArrgError is not in training. 
For MBI, Parameter Matching decreases from $92\%$ to $77\%$ when it is excluded with Resource Leak.
Our model is unable to detect Epoch Lifecycle (accuracy of 0) if Parameter Matching, Call Ordering or Message Race is also removed from training. Like MissplacedCall, Message Race has a higher accuracy if Parameter Matching is removed from training. 

In other words, these prediction scores can be used to quantify how much two errors share the same code patterns:  MissingCall from MPI-CorrBench has similar code patterns as ArgError. We believe such metrics have potential to guide the errors topology definition. 

\begin{table*}
\centering%
\begin{tabular}{cc|ccccccccc}
\toprule
&& \multicolumn{6}{c}{\textbf{Compilation option - code correct/incorrect}}\\
\textbf{Training Dataset}& \textbf{Features}  & \textit{O0-ok} & \textit{O2-ok} & \textit{Os-ok}& \textit{O0-ko} & \textit{O2-ko} & \textit{Os-ko}\\
\midrule
MBI & all & \cellcolor{green!30}ok& \cellcolor{green!30}ok& \cellcolor{green!30}ok& \cellcolor{red!30}ok& \cellcolor{red!30}ok& \cellcolor{red!30}ok\\
MPI-CorrBench & all & \cellcolor{green!30}ok& \cellcolor{green!30}ok& \cellcolor{green!30}ok& \cellcolor{red!30}ok& \cellcolor{red!30}ok& \cellcolor{red!30}ok\\
MBI & GA & \cellcolor{red!30}ko& \cellcolor{green!30}ok& \cellcolor{green!30}ok& \cellcolor{green!30}ko& \cellcolor{green!30}ko& \cellcolor{green!30}ko&\\
MPI-CorrBench & GA  & \cellcolor{red!30}ko& \cellcolor{green!30}ok& \cellcolor{green!30}ok& \cellcolor{green!30}ko& \cellcolor{green!30}ko& \cellcolor{green!30}ko&\\

\bottomrule
\end{tabular}
\vspace{.1cm}
\caption{Prediction on Hypre using models trained on either MBI or MPI-CorrBench. \textit{ok} refers to correct codes while \textit{ko} refers to incorrect codes. Each column on the right side represents a version of Hypre (either correct or incorrect) compiled with \textit{-O0}, \textit{-O2}, or \textit{-Os}. Each line represents a model trained using the dataset along with the described features. 
The value of the cell indicates the model prediction on the column code: ok or ko for correct or incorrect, respectively. The cell color shows if the model correctly predicted (in green) the code label or if it made an error (red).}\label{tab:real}
\end{table*}


\subsection{Preliminary Real Case Scenario}

A limitation of our approach is the scale of the experiments. In our knowledge, MPI-CorrBench and MBI are the only two benchmarks with correct and incorrect MPI codes. We can use mutation techniques or GitHub to acquire new incorrect cases, but we decided to start with these existing correctness benchmarks as they are the standards for evaluating active verification tools, enabling a fair comparison. Large-scale exploration is a promising future direction for us that we discuss in Section~\ref{sec:lim}. 

To try our models beyond benchmarks, on a real case, we investigate Hypre, a library of high performance preconditioners and solvers featuring multigrid methods, available on Github \footnote{\url{https://github.com/hypre-space/hypre}}. An error due to the use of the same tag in two MPI operations is fixed on commit bc3158e (version 2.10.1). We retrieved the code before and after this commit in order to have a correct and incorrect version of the code. Both versions of the code can be used by our models to evaluate if the cross modeling has the potential to detect errors in real world applications.



To extract the features, we start by compiling the codes. As described previously, 
we consider the tree compiler options: \textit{-O0}, \textit{-O2}, and \textit{-Os}. Each resulting IR is used by IR2Vec to generate vectors representing the codes. The vectors are subsequently normalized with \textit{vector} as in the \textit{IR2Vec Cross} evaluation. We trained our models on either MBI or MPI-CorrBench. 

Table~\ref{tab:real} presents the predictions of the different models. We evaluated the models without feature selection (\textit{all}), or with GA (\textit{GA}) following the same procedure as in Section~\ref{subsec:intra}. Without features selection, our models fail to detect the error in Hypre (red parts in the first two lines). However, when we select different features, the models successfully label the code (and correctly predict the \textit{cross} scenario), independently if they were trained on MBI or MPI-CorrBench. We note that by changing the features, \textit{O0-ok} can accurately be predicted as correct, but that we did not find any combination of features that successfully label all Hypre versions. This further indicates that the compiler optimization used to generate the code representation must be considered with the machine learning models as both impact the predictions.

\section{Discussion and Future Work}
\label{sec:lim}

Detecting errors in MPI programs is challenging, and no method is able to detect all kinds of errors. Expert tools and ML-based approaches try to address this challenge and seem to achieve similar results. Yet, they employ drastically different approaches. Expert tools require human expertise to identify and manually devise algorithms and heuristics implemented in the tool for diverse error patterns. While costly in manpower and time, such approaches could provide the benefit of understanding why a code is assumed incorrect. Conversely, with ML methods, currently, we cannot easily understand why a code is predicted as incorrect (or ensure that a predicted code is correct for the errors that our model predicts). This is a limitation of ML strategies, which is now an important research direction in explainable AI research. Nevertheless, ML methods only require a new dataset to consider new bug issues. Thus, we expect such methods to easily generalize over new emerging scenarios.

Indeed, and as future work, we plan to apply our models on larger scales. By crawling GitHub repositories, we can use our models as detectors to identify bugs in existing MPI projects. We can also take the GitHub codes as additional training datasets for our models. The main challenge will be the labeling of the data. Our insights are to track how specific errors impact code embedding or look at the GitHub metadata. We also consider training models to predict the error location. A first step in that direction is applying our models at different code granularities by extracting the code into different compilation units. Whether or not an error is detected across the different compilation units can serve as a guideline for the exact error location and what caused it. Finally, we envision training Large Language Models to propose code fixes directly. 

\section{Conclusion}
\label{sec:conclusion}

To the best of our knowledge, this paper presents the first method using machine learning techniques to detect errors in MPI programs. We developed two models that either use embedding or deep learning graph neural networks. We trained and validated these models on three datasets and compared them with existing MPI verification tools. The ML methods achieved competitive results across the datasets compared to the expert tools. Furthermore, while our models do not provide feedback like expert specialized tools, they show promising generalization capabilities over new unseen error types or benchmark suites. These results show the potential of ML-based approaches in the context of MPI verification. 

\section*{Acknowledgments}
\label{sec:ack}

The authors would like to thank Philippe Virouleau for evaluating the verification tool ITAC over MBI. 
We would also like to thank the Research IT team\footnote{https://researchit.las.iastate.edu} of Iowa State University for their continuous support in providing access to HPC clusters for conducting the experiments of this research project. This project was also supported by the National Science Foundation (NSF) under grant number 2211982. Experiments presented in this paper were carried out using the experimental testbeds PlaFRIM, supported by Inria, CNRS (LABRI and IMB), Université de Bordeaux, Bordeaux INP, Conseil Régional d’Aquitaine\footnote{https://www.plafrim.fr}, and Grid'5000, supported by a scientific interest group hosted by Inria, 
CNRS, RENATER and several Universities as well as other organizations\footnote{https://www.grid5000.fr}.
Finally, we would like to thank the anonymous reviewers for their feedback.


\bibliographystyle{IEEEtran}
\bibliography{biblio}

\end{document}